\documentstyle[12pt,twoside]{article}
\def\greaterthansquiggle{\raise.3ex\hbox{$>$\kern-
.75em\lower1ex\hbox{$\sim$}}}
\def\lessthansquiggle{\raise.3ex\hbox{$<$\kern-
.75em\lower1ex\hbox{$\sim$}}}
\newcommand{\beq}{\begin{equation}}
\newcommand{\eeq}{\end{equation}}
\newcommand{\beqa}{\begin{eqnarray}}
\newcommand{\eeqa}{\end{eqnarray}}
\newcommand{\beqan}{\begin{eqnarray*}}
\newcommand{\eeqan}{\end{eqnarray*}}
\newcommand{\ba}{\begin{array}}
\newcommand{\ea}{\end{array}}
\newcommand{\no}{\nonumber}

\newcommand{\A}{{\cal A}}

\newcommand{\C}{{\cal C}}

\newcommand{\F}{{\cal F}}

\def\nz{\ifmmode {I\hskip -3pt N} \else {\hbox {$I\hskip -3pt N$}}\fi}
\def\zz{\ifmmode {Z\hskip -4.8pt Z} \else
       {\hbox {$Z\hskip -4.8pt Z$}}\fi}
\def\qz{\ifmmode {Q\hskip -5.0pt\vrule height6.0pt depth 0pt
       \hskip 6pt} \else {\hbox
       {$Q\hskip -5.0pt\vrule height6.0pt depth 0pt\hskip 6pt$}}\fi}
\def\rz{\ifmmode {I\hskip -3pt R} \else {\hbox {$I\hskip -3pt R$}}\fi}
\def\cz{\ifmmode {C\hskip -4.8pt\vrule height5.8pt\hskip 6.3pt} \else
       {\hbox {$C\hskip -4.8pt\vrule height5.8pt\hskip 6.3pt$}}\fi}
\def\au{{\setbox0=\hbox{\lower1.36775ex%
\hbox{''}\kern-.05em}\dp0=.36775ex\hskip0pt\box0}}
\def\ao{{}\kern-.10em\hbox{``}}

\renewcommand{\baselinestretch}{1.5} 
\normalsize
\begin{document}
\bibliographystyle{plain}

\begin{titlepage}
\begin{flushright}
UWThPh-2002-15 \\

\end{flushright}
\vspace{2cm}
\begin{center}
{\Large \bf  BV-Quantization} \\\ {\Large \bf of a} \\\
{\Large \bf  Noncommutative Yang--Mills 
Theory
Toy Model}\\[40pt]
$ \mbox{Helmuth H\"uffel}^{1,2}$\\
Institut f\"ur Theoretische Physik \\
Universit\"at Wien \\
Boltzmanngasse 5, A-1090 Vienna, Austria
\vfill

{\bf Abstract}
\end{center}
\renewcommand{\baselinestretch}{1.0} 
\small
We review the Batalin-Vilkovisky quantization 
procedure for Yang--Mills theory on a 2-point space.

\vfill
\begin{enumerate}
\item[1)] Talk given at the 5-th 
International Conference 
"Renormalization Group 2002", Tatranska Strba, Slovakia, 
March 10-16, 2002
\item[2)] Email: helmuth.hueffel@univie.ac.at
\end{enumerate}
\end{titlepage}
\renewcommand{\baselinestretch}{1.5} 
\normalsize

\setcounter{equation}{0}

In this talk we give a short summary of \cite{h},
where we proposed the quantization of 
one of the simplest toy 
 models for noncommutative gauge theories  which is (zero dimensional)
Yang--Mills theory 
 on a 2-point space. 
 
Noncommutative geometry constitutes one of the  fascinating new concepts 
in current theoretical physics research with many promising applications
\cite{Connes,Mad,Landi,GracVar,Kastler}.

We quantize the Yang--Mills theory on a 2-point space by applying the
standard 
Batalin--Vilkovisky   method \cite{BV,BV2}.
Somewhat surprisingly we find that despite of the model's  original
simplicity 
the gauge structure 
reveals infinite reducibility 
and the gauge fixing is afflicted with the Gribov \cite{Gribov} problem. 

The basic idea of noncommutative 
 geometry is to replace the notion of  differential manifolds  and
functions 
 by specific    
 noncommutative  
 algebras of
 functions.
Following \cite{ConnL} we define the Yang--Mills Theory on a 2-point space
in terms of the algebra ${\mathbf A}=C \oplus C$ which is represented 
 by diagonal  complex valued $2\times 2$ 
 matrices. 
The differential p-forms are constant, diagonal or 
offdiagonal  $2\times 2$  matrices, 
depending on whether p is even or odd,
respectively. A  nilpotent derivation $\bf d$ acting on   
$2\times 2$ matrices is defined by
 ${\bf d}\,a =i \left( \ba{cc}
a_{21}+a_{12} &  a_{22}-a_{11}  \\
a_{11}-a_{22} & a_{21} +a_{12}
\ea \right) 
\qquad {\rm where } \qquad
a = \left( \ba{cc}
a_{11} & a_{12}  \\
a_{21} & a_{22} 
\ea \right), \quad a_{ij} \in \rm C. 
$
The anti-Hermitean 1-forms $\A$  can be 
parametrized by
\begin{equation}
\A = \left( \ba{cc}
0 & i  \phi  \\
i  \bar \phi  & 0 
\ea \right) 
\end{equation}
and constitute the gauge 
fields of the model; here $\phi\in \rm C$  
denotes a 
(constant)
scalar field.
The (rigid) gauge 
transformations of $\A$ are defined by
\begin{equation}
\label{old}
\A^{U}=U^{-1}\A U+U^{-1}{\bf d}\,U
\end{equation}
with $U$ being  a unitary element of the algebra $\mathbf{A}$. It is a 
constant, 
 diagonal  and unitary matrix which can  be 
parametrized by the diagonal   matrix 
$\varepsilon$
\begin{equation}
U = \left( \ba{cc}
 e^{i \alpha} & 0  \\
0 & e^{i \beta} 
\ea \right)=e^{i \varepsilon}, 
\quad \varepsilon=\left( \ba{cc}
\alpha & 0  \\
0 & \beta 
\ea \right)\quad\alpha,\beta \in \rm R.
\end{equation}
Due to the nonabelian form of the gauge transformations 
(\ref{old}) the $U(1)\times U(1)$ gauge model shares many 
interesting features with the standard  Yang--Mills 
theory, yet   it has no physical 
space-time dependence and allows  extremely simple  calculations.

We define a 
scalar product for $2\times 2$ matrices $a,b$ by $\langle 
a \,\vert\, b \rangle=tr\, a^{\dagger} \,b$ where $\dagger$ denotes
taking the  Hermitian conjugate. 
The curvature $\F$ is defined as usual by $\F={\bf d}\A+\A\,\A$ and
for an action which is  automatically invariant  under the gauge 
transformations (\ref{old})  one takes
\begin{equation}
\label{action}
S_{inv}=\frac{1}{2}\langle \F \vert \F \rangle=
\left( (\phi+\bar\phi)+\phi \,\bar \phi\right)^2.
\end{equation}

To discuss  infinitesimal  (zero-stage) gauge transformations we
introduce 
a  diagonal infinitesimal (zero stage) gauge parameter matrix 
$\varepsilon^0_{e}$ in terms of which
 $U \simeq {\bf 
1}+\varepsilon^0_{e}$. The  infinitesimal (zero-stage) gauge variation
of $\A$
derives as 
\begin{equation}
\label{zero}
\delta_{\varepsilon^0_{e}}{\A}=i {\bf R}^0 \,\varepsilon^0_{e} \quad 
{\rm where} \quad
{\bf R}^0={\bf D};
\end{equation}
here the \mbox{(zero-stage)} gauge generator ${\bf R}^0$ is defined in
terms of the
 covariant matrix derivative ${\bf D}$, which acting on 
$\varepsilon^0_{e}$ is given by 
${\bf D}\varepsilon^0_{e} ={\bf 
d}\varepsilon^0_{e}+[\A,\varepsilon^0_{e}]$.

A gauge symmetry is called  irreducible if the  (zero 
stage) gauge generator ${\bf 
R}^0$ does not possess any zero mode. It is amusing to note that the
Yang--Mills theory on the 2-point 
space reveals an infinitely reducible gauge symmetry: We observe 
that
  ${\bf D\,d}$ is vanishing on arbitrary offdiagonal matrices. Thus
there 
exists a zero mode $\varepsilon^1_{e}$ for  the (zero-stage) gauge 
generator 
${\bf R}^0$, such that
\begin{equation}
{\bf R}^0\varepsilon^1_{e}=0 \quad{\rm where} \quad
\varepsilon^1_{e}={\bf R}^1
\varepsilon^1_{o} \quad {\rm with} \quad{\bf R}^1={\bf d}.
\end{equation}
Here $\varepsilon^1_{o}$ denotes    an offdiagonal, infinitesimal
(first-stage) 
gauge parameter matrix and
  ${\bf R}^1$ the corresponding (first-stage) gauge generator.
As a matter of fact  
an infinite tower of  (higher-stage) gauge generators
${\bf R}^s$,\,\,\mbox{$s=1,2,3,\,\cdots$} with never ending
   gauge invariances for 
gauge invariances is arising: We define  ${\bf R}^s={\bf d}$
 for \mbox{$s=1,2,3,\,\cdots$} 
so that  for each  gauge generator
 there exists an 
additional zero mode 
\begin{eqnarray}
{\bf R}^1\varepsilon^2_{o}=0, \quad &{\rm 
where}&\quad\varepsilon^2_{o}={\bf R}^2\varepsilon^2_{e} \no \\
{\bf R}^2\varepsilon^3_{e}=0, \quad &{\rm 
where}&\quad\varepsilon^3_{e}={\bf R}^3\varepsilon^3_{o}\no\\
\cdots\qquad&   &\qquad\cdots
\end{eqnarray}
due to the nilpotency ${\bf d}^2=0$.

Now we straightforwardly apply the 
usual field theory 
BV-path integral quantization scheme \cite{BV,BV2}  
to the  2-point model:  
In addition to the original gauge field  $\A\equiv\C^{-1}_{-1}$ we
introduce  
  ghost fields 
$\C^{k}_s$, \, $\infty\ge s\ge -1,\,\,s\ge k\ge -1$ with $k \,\,{\rm
odd}$, as well 
as auxiliary ghost fields 
$\bar{\C}^{k}_s$, \, $\infty\ge s\ge 0,\,\,s\ge k\ge 0$ with $k \,\,
{\rm 
even}$.
Furthermore we add Lagrange multiplier fields $\pi^{k}_s$, 
 $\infty\ge s\ge 1,\,s\ge k\ge 1$ with $k \,\,{\rm odd}$ and
$\bar{\pi}^{k}_s$,  $\infty\ge s\ge 0,\,s\ge k\ge 0$ with $k \,\, {\rm 
even}$; finally we introduce antifields ${\C^{k}_s}^{*}$, ${\bar{\C}^{k}_s\,}^{*}$.
The BV-action  obtains  as
\begin{eqnarray}
S_{BV}&=&S_{inv}+S_{aux}-\langle{{\cal C}_{-1}^{-1}}^{*} \vert{\bf 
D}\,{\cal 
C}_{0}^{-1}\rangle -\sum_{s=1,3,5, \cdots}^{\infty} \langle{{\cal
C}_{s}^{-1}}^{*} 
\vert{\bf d}\,{\cal 
C}_{s+1}^{-1}\rangle
\no\\
&&\mbox{} 
\qquad\qquad\qquad\qquad\qquad\quad \,\,\,\,\,
-i\sum_{s=0,2,4, \cdots}^{\infty} \langle{{\cal C}_{s}^{-1}}^{*} 
\vert{\bf d}\,{\cal 
C}_{s+1}^{-1}\rangle,
\end{eqnarray}
where we denote by $S_{aux}$ the auxiliary field action 
\beq
S_{aux}=\sum_{k=0,2,4, \cdots}^{\infty} \,\sum_{s=k}^{\infty} 
\langle{\bar\pi}^{k}_{s}  \vert
{{{\bar{\cal C}}_{s}^{k}}\,}^{*}\rangle+
\sum_{k=1,3,5, \cdots}^{\infty} \,\sum_{s=k}^{\infty} 
\langle{{\C}^{k}_{s}}^{*} \vert
{{{{\pi}}_{s}^{k}}\,}\rangle.
\end{equation}
Gauge fixing conditions  similar to the usual Feynman gauge  are
implemented  by introducing
 the gauge fixing fermion 
$\Psi=\Psi_{\delta}+\Psi_{\pi}$ 
\begin{eqnarray}
\Psi_{\delta}
&=&
\sum_{s=0,2,4, \cdots}^{\infty}\,\,\sum_{k=0,2,4, \cdots\, \, k\le s} \, 
\left(-\langle{\bar\C}^{k}_{s} \,\vert\,
{{{\mbox{\boldmath $\delta$}{\cal C}}_{s-1}^{k-1}}\,}\rangle +
\langle{{\mbox{\boldmath $\delta$}\bar\C}^{k}_{s+1} \,\vert\,
{{{\cal C}}_{s+2}^{k+1}}\,}\rangle 
\right.\no\\
&&\mbox{} 
\qquad\qquad\qquad\qquad\quad+
\left. 
i\langle{\bar\C}^{k}_{s+1} \,\vert\,
{{{\mbox{\boldmath $\delta$}{\cal C}}_{s}^{k-1}}\,}\rangle +
i\langle{{\mbox{\boldmath $\delta$}\bar\C}^{k}_{s} \,\vert\,
{{{\cal C}}_{s+1}^{k+1}}\,}\rangle\right),\no \\
\Psi_{\pi}&=&
\frac{1}{2} \sum_{s=0,2,4, \cdots}^{\infty}\,\,\sum_{k=0,2,4, \cdots\, 
\, k <  s} 
\left(\langle{\bar\C}^{k}_{s} \,\vert\,
{{{{\pi}}_{s}^{k+1}}\,}\rangle+
\,\langle{\bar\pi}^{k}_{s} \,\vert\,
{{{{\C}}_{s}^{k+1}}\,}\rangle
\right.\no\\
&&\mbox{} +
\left. 
i\langle{\bar\C}^{k}_{s+1} \,\vert\,
{{{{\pi}}_{s+1}^{k+1}}\,}\rangle+
\,i\langle{\bar\pi}^{k}_{s+1} \,\vert\,
{{{{\C}}_{s+1}^{k+1}}\,}\rangle
\right)+ \frac{1}{2}
\sum_{k=0,2,4, \cdots}^{\infty} \, 
\langle{\bar\C}^{k}_{k} \,\vert\,
{{{\bar{\pi}}_{k}^{k}}\,}\rangle.
\end{eqnarray}
By $\mbox{\boldmath $\delta$}$  we denote a nilpotent
 coderivative operator  
$\mbox{\boldmath $\delta$} \, a =$ $i \left( \ba{cc}
a_{12}-a_{21} & -a_{11}-a_{22}  \\
-a_{11}-a_{22} & -a_{12}+a_{21}
\ea \right)$  where 
$a = \left( \ba{cc}
a_{11} & a_{12}  \\
a_{21} & a_{22} 
\ea \right)$,  $a_{ij} \in \rm C$.  
We eliminate the
antifields  by using the gauge fixing fermion 
$\Psi$ via
\beq
\langle{\C^{k}_s}^{*}\vert=\frac{\partial\Psi}{\partial \vert 
\C^{k}_s\rangle}, \quad
\vert{\bar\C^{k}_s\,}^{*}\rangle=\frac{\partial \Psi}{\partial 
\langle\bar\C^{k}_s\vert}, 
\end{equation}
so that the gauge fixed action $S_{\Psi}$ reads
\begin{eqnarray}
S_{\Psi}&=&S_{inv}- i \langle {\bar \C}_{0}^{0} \,\vert\,
\mbox{\boldmath $\delta$} 
{\bf D}\,
{\C}_{0}^{-1}\rangle
 \no\\
&&\mbox{}
-i
\sum_{s=1,3,5, \cdots}^{\infty} \langle{\bar \C}_{s+1}^{0}
 \,\vert\, \mbox{\boldmath $\delta$}{\bf d}\,{\C}_{s+1}^{-1}\rangle +
 \sum_{s=0,2,4, \cdots}^{\infty} \langle{\bar \C}_{s+1}^{0}
 \,\vert\, \mbox{\boldmath $\delta$}{\bf d}\,{\C}_{s+1}^{-1}\rangle 
 \no\\
&&\mbox{}+ 
\sum_{k=0,2,4, \cdots}^{\infty} \,\, \sum_{s=k+1,\, odd}^{\infty}
 \left(i \langle{{{{\bar\pi}}_{s}^{k}}} \,\vert\, 
{\pi}^{k+1}_{s}\rangle 
+\langle{\bar\pi}^{k}_{s}  \, 
\vert \,(i  \mbox{\boldmath $\delta$}{{{{\C}}_{s-1}^{k-1}}\,}+
 \,{\bf d}{\C}^{k+1}_{s+1})\rangle\no \right.\\
&&\mbox{}\left. \qquad\qquad\qquad\qquad\quad+\langle(i
\mbox{\boldmath $\delta$}{{{{\bar\C}}_{s-1}^{k}}\,}\,-
{\bf d}{\bar\C}^{k+2}_{s+1})\vert\,  
{{\pi}}_{s}^{k+1}\rangle\right)\no\\
&&\mbox{}+ 
\sum_{k=0,2,4, \cdots}^{\infty}\,\,  \sum_{s=k+2,\, even}^{\infty}
 \left( \langle{{{{\bar\pi}}_{s}^{k}}} \,\vert\, 
{\pi}^{k+1}_{s}\rangle 
+\langle{\bar\pi}^{k}_{s}  \, 
\vert \,(-  \mbox{\boldmath $\delta$}{{{{\C}}_{s-1}^{k-1}}\,}+
 \,i{\bf d}{\C}^{k+1}_{s+1})\rangle \right. \no\\
&&\mbox{}\left. \qquad\qquad\qquad\qquad\quad+\langle(
\mbox{\boldmath $\delta$}{{{{\bar\C}}_{s-1}^{k}}\,}\,+
i{\bf d}{\bar\C}^{k+2}_{s+1})\vert\,  
{{\pi}}_{s}^{k+1}\rangle\right)\no\\
&&\mbox{}+\sum_{k=0,2,4, \cdots}^{\infty} \,
\langle{\bar\pi}^{k}_{k}  \,\vert\, 
( - \mbox{\boldmath $\delta$}{{{{\C}}_{k-1}^{k-1}}\,}+
i{\bf d}{\C}^{k+1}_{k+1} \,+\frac{1}{2}
{{{{\bar\pi}}_{k}^{k}}})\rangle. 
\end{eqnarray}
We can now  eliminate the Lagrange multiplier fields $\pi^{k}_s$ 
and $\bar{\pi}^{k}_s$  and  arrive at
\begin{eqnarray}
S_{\Psi}\longrightarrow &&S_{inv}+ \frac{1}{2}\langle\A \,\vert \,
{\bf d}\mbox{\boldmath $\delta$} 
\,
\A\rangle-i
\langle{\bar \C}_{0}^{0}\,\vert\,(
\mbox{\boldmath $\delta$} 
{\bf D}\,+{\bf d}\mbox{\boldmath $\delta$})\,
{\C}_{0}^{-1}\rangle \no\\
&&\mbox{}-i\sum_{s=1,3,5, \cdots}^{\infty} \langle{\bar \C}_{s+1}^{0}
 \,\vert\, (\mbox{\boldmath $\delta$}{\bf d}+{\bf d}\mbox{\boldmath 
 $\delta$})\,{\C}_{s+1}^{-1}\rangle\no\\
&&\mbox{} +\sum_{s=0,2,4, \cdots}^{\infty} \langle{\bar \C}_{s+1}^{0}
 \,\vert\, (\mbox{\boldmath $\delta$}{\bf d}+{\bf d}\mbox{\boldmath 
 $\delta$})\,{\C}_{s+1}^{-1}\rangle \no\\
&&\mbox{}-
i\sum_{k=0,2,4, \cdots}^{\infty}\, \,\sum_{s=k+1, \, odd}^{\infty}
\langle{\bar \C}_{s+1}^{k+2}\,\vert\,(
\mbox{\boldmath $\delta$} 
{\bf d}\,+{\bf d}\mbox{\boldmath $\delta$})\,
{\C}_{s+1}^{k+1}\rangle\no\\
&&\mbox{}+
\sum_{k=0,2,4, \cdots}^{\infty}\, \,\sum_{s=k+2, \, even}^{\infty}
\langle{\bar \C}_{s+1}^{k+2}\,\vert\,(
\mbox{\boldmath $\delta$} 
{\bf d}\,+{\bf d}\mbox{\boldmath $\delta$})\,
{\C}_{s+1}^{k+1}\rangle\no\\
&&\mbox{} +\frac{1}{2}\sum_{k=0,2,4, \cdots}^{\infty}  
\langle
{\C}_{k+1}^{k+1}\,\vert\,(
\mbox{\boldmath $\delta$} 
{\bf d}\,+{\bf d}\mbox{\boldmath $\delta$})\,
{\C}_{k+1}^{k+1}\rangle 
.
\end{eqnarray}
All the higher-stage ghost  contributions  can be integrated away 
 as
$\mbox{\boldmath $\delta$} 
{\bf d}\,+{\bf d}\mbox{\boldmath $\delta$}=4 \cdot{\bf 1}$ and we 
simply
obtain
\beq
\label{fixfinal}
S_{\Psi} \longrightarrow S_{inv}+ \frac{1}{2}\langle\A \,\vert \,
{\bf d}\mbox{\boldmath $\delta$} 
\,
\A\rangle-
i\langle{\bar \C}_{0}^{0}\,\vert\,(
\mbox{\boldmath $\delta$} 
{\bf D}\,+{\bf d}\mbox{\boldmath $\delta$})\,
{\C}_{0}^{-1}\rangle.
\end{equation}

We summarize that the zero dimensional Yang--Mills theory model on a 
 2-point space reveals infinite reducibility; 
after applying the standard BV-quantization procedure
the action finally  contains invertible quadratic parts 
for the gauge field, as well as
for  the ghost fields. A closer inspection \cite{h} shows that the 
model  suffers from a Gribov 
problem \cite{Gribov}.

We expect that our present investigations  
will lead to a   study of the renormalization 
effects at higher orders; it may also be possible to  
compare the  perturbative calculations 
with  explicit analytic 
integrations (for related attempts see \cite{Hauss}).

\section*{Acknowledgements}

The author wishes to thank the organizers for their kind 
invitation to this conference.

\end{document}